\documentclass[aps,preprint,floatfix]{revtex4-1}
\usepackage[utf8x]{inputenc}
\usepackage{amsmath}
\usepackage{graphicx}

\draft % marks overfull lines with a black rule on the right

\begin{document}

\renewcommand{\baselinestretch}{1.50}\normalsize




%\preprint{}

\title{Assessment of hydrocarbon electron-impact ionization cross section measurements 
for magnetic fusion} %Title of paper








\author{Stefan E. Huber}

\email[]{Author to whom correspondence should be addressed. Electronic mail: s.huber@uibk.ac.at}

\author{Josef Seebacher}

\author{Alexander Kendl}


%\thanks{}


\affiliation{Institut f\"ur Ionenphysik und Angewandte Physik, Universit\"at Innsbruck, Assoziation 
EURATOM-\"OAW, Technikerstr. 25, A-6020 Innsbruck, Austria}

\author{Detlev Reiter}

\affiliation{Institut f\"ur Energieforschung - Plasmaphysik Forschungszentrum J\"ulich GmbH, EURATOM 
Association, Trilateral Euregio Cluster, D-52425 J\"ulich, Germany}





\date{May 29, 2011}

\begin{abstract}
Partial ionization cross section experiments have been carried 
out recently at the University of Innsbruck for three types of hydrocarbons, i.e. acetylene, ethylene 
and propene. 
Cross section data fits are generated and compared to the compilation of earlier experimental data summarized 
in the online database HYDKIN [www.hydkin.de]. New data fits are brought into a suitable form to be 
incorporated into the 
database. 
In order to illuminate underlying dissociation mechanisms the energy dependence of branching 
ratios above energies of $20-30eV$ is reviewed in light of the present results. \\
This is a pre-peer reviewed version which has been submitted to Contributions to Plasma Physics.
\end{abstract}


\maketitle %\maketitle must follow title, authors, abstract and \pacs



\section{Introduction}
\label{intro}
Since graphite is still a candidate as wall material for the high flux zones at the divertor of the 
fusion experiment ITER, hydrocarbon impurities will be formed due to chemical erosion \cite{samm,ph,fed}. A 
key ingredient for the simulation of the transport, chemistry, and radiation behaviour of these impurities 
are the cross sections for collision processes with 
electrons and protons present in the scrape-off layer (SOL) plasma \cite{jrethanpropanpaper,jrmethanpaper}. 
For that reason, the HYDKIN cross section database 
\cite{reiterkueppers} has been set up over the last decades to cover the information on such cross sections. 
Much experimental cross sectional data have been compiled and a revision of many of the data used in HYDKIN 
has recently been carried out and presents the current state of the HYDKIN database \cite{jrcpp}.Concerning 
especially (dissociative) ionization cross sections, fit curves of experimental data have been determined by 
making use of the following fitting expression \cite{jrethanpropanpaper,jrmethanpaper,jrcpp}:
\begin{equation}
\label{fitformula}
 \sigma(E) = \frac{10^{-13}}{E E_{th}}\left[A_1\ ln(E/E_{th}) + \sum^N_{j=2} A_j 
\left(1-\frac{E_{th}}{E}\right)^{j-1}\right],
\end{equation}
where $\sigma(E)$ denotes the cross section in units of $cm^2$, $E$ is the collision energy expressed in 
$eV$, and $A_j, j \in \{2,\dots,N\}$ are fitting parameters. $N$ has been set to values such that the r.m.s. 
of the fit is smaller than $2-3\%$, i.e. $N=6$ in most cases. The factor $10^{-13}$ has been singled out 
from the fitting parameters in order to make them more handy for both the user and the used fit software. 
$E_{th}$ is the appearance potential (also expressed in $eV$), depending on the considered process. In 
addition, cross sections have been generated for processes, for which no experimental data have been 
available, by making use of well-based (auxiliary) assumptions, like energy invariance of branching ratios 
(see Sec. \ref{invariance}), and certain well established cross section scaling rules 
\cite{jrethanpropanpaper,jrmethanpaper,jrcpp}. \\
In 2006 and 2009, respectively, partial electron impact ionization cross section experiments have been 
carried out at the University of Innsbruck for three types of hydrocarbons, i.e. acetylene \cite{feil1}, 
ethylene \cite{endstrasser} and propene \cite{feil2}. In this work the results of these measurements are first 
used to generate cross section fits for the processes covered by these experiments to be subsequently 
compared to the compilation of experimental data available in HYDKIN for acetylene \cite{c2h2hydkin}, 
ethylene \cite{c2h4hydkin} and propene \cite{c3h6hydkin1,c3h6hydkin2}. It should be kept in mind that when 
the discussion appears to be about agreement of 
new cross section fits with the HYDKIN database, this is just short for a comparison of present 
experimental data to earlier experimental data on which the HYDKIN database is based.\\
After summarizing shortly the methodology of fitting the recent experimental data as well as the sources 
for ionization and appearance potentials in Sec. \ref{method}, the new cross section fits are presented and 
compared to already existing cross section expressions provided by HYDKIN in the subsequent Secs. 
\ref{acetylene}-\ref{propene}. In Sec. \ref{invariance} the energy dependence of branching ratios for large 
energies is reviewed in light of the recent data. Finally, in Sec. \ref{summary}, a conclusion is summarized.

\section{Methodology}
\label{method}
In order to generate cross section fits considering partial electron impact ionization cross sectional data, 
expression \ref{fitformula} has been used to determine the fitting coefficients $A_j$ with $N=6$ in most 
cases. It has been shown elsewhere that cross sections given by Eq. (\ref{fitformula}) provide a proper 
physical energy dependence in the threshold and high energy regions with the fitting coefficients $A_j$ 
fulfilling certain conditions \cite{jrcpp}. These conditions are met by the coefficients for the processes 
presented in Secs. \ref{acetylene}-\ref{propene}. In case of electron impact cross sections the collision 
energy $E$ can be set approximately equal to the electron impact energy with high 
accuracy due to the small mass of the electron compared the molecular masses of the molecules in 
consideration. The value of $E_{th}$ is given by the ionization potential in case of ordinary ionization 
cross sections, and by the appearance potential in case of dissociative ionization processes, where the 
(experimental) appearance energy is defined as the experimentally accessible minimum energy value for the 
appearance of ionic fragments \cite{ill}. The numerical values for both the ionization and the 
appearance potentials for the processes discussed in this work have mainly been adopted from Refs. 
\cite{feil1,locht,davister1,davister2} concerning acetylene, from Refs. \cite{feil2,plessis} concerning 
ethylene, and from the data compilation provided by NIST chemistry webbook \cite{nist} concerning propene as 
well as some specific dissociation channels concerning the former two hydrocarbons. 
In specific cases no experimental values for the threshold energy $E_{th}$ have been available. The choice of 
$E_{th}$ concerning such cases will be discussed when discussing the cross section fits for the corresponding 
processes in Secs. \ref{acetylene}-\ref{propene}.

\section{Results}
\subsection{Acetylene}
\label{acetylene}
Ionization cross section fits for the following electron-impact ionization processes of acetylene, revealed 
by the measurements of Feil et al. \cite{feil1}, have been generated:
\begin{subequations}
 \begin{eqnarray}
  e + C_2H_2 & \rightarrow & C_2H_2^+ + 2e, \label{c2h21} \\
   & \rightarrow & C_2H^+ + H + 2e, \label{c2h22} \\
   & \rightarrow & C_2^+ + \dots, \label{c2h23} \\
   & \rightarrow & CH^+ + \dots, \label{c2h24} \\
   & \rightarrow & C^+ + \dots, \label{c2h25} \\
   & \rightarrow & CH_2^+ + C^+ + 3e, \label{c2h26} \\
   & \rightarrow & C_2H_2^{2+} + 3e, \label{c2h27}
 \end{eqnarray}
\end{subequations}
where the dots designate the sum of all possible dissociation sub channels. 
Since it is only possible to distinguish product ions with a mass spectrometer with respect to their mass to 
charge ratio, in a first step a cross section for the sum of processes (\ref{c2h25},\ref{c2h27}) has been 
measured by Feil et al. \cite{feil1}, because both $C_2H_2^{2+}$ and $CH^+$ have mass to charge ratio of $13$ 
Thompson. In a second step cross sections differential with respect to ion kinetic energy have been 
determined. These yield cross sections for fragment ions with low initial kinetic energy ($<0.5eV$) 
and high initial kinetic energy ($0.55-10eV$). As ionization of $C_2H_2$ resulting in $C_2H_2^{2+}$ does not 
change the momentum of the product ion, the low kinetic energy part has been assigned to the formation of 
$C_2H_2^{2+}$, whereas the high energy part has been assigned to the production  of $CH^+$ fragments 
\cite{feil1}. In addition, also the threshold of the low kinetic energy cross section has been observed to be 
about $36eV$ which corresponds nicely with the ionization energy of $^{13}C^{12}CH_2^{2+}$, which has been 
determined separately \cite{feil1}. \\
The fitting coefficients $A_j$, according to Eq. (\ref{fitformula}) as well as the threshold energies 
$E_{th}$ for the cross section fits for the processes listed above are given in table \ref{c2h2table}.

\begin{table}[h]
\renewcommand{\baselinestretch}{1}\normalsize
\caption{\label{c2h2table}Values of fitting parameters $E_{th}$ (with references in squared brackets) and 
            $A_i$ in Eq. (\ref{fitformula}) for partial ionization cross sections of acetylene.}
\begin{tabular}{@{}lcccc@{}}
\hline
 Reaction $e + C_2H_2 \rightarrow$ & $E_{th}$ & $A_i, i=1-3$ &  \\
	 &	         & $A_i, i=4-6$ &  \\
\hline
 $C_2H_2^+ + 2e$ & 1.14000E+01 \cite{reiterkueppers,nist} & 3.73243E+00 & -3.73243E+00 & -8.01475E-01 \\
				   &	   & 3.15344E+00 & -1.00779E+01 & 7.28853E+00 \\
 $C_2H^+ + H + 2e$ & 1.73000E+01 \cite{davister2} & 7.96959E-02 & -7.96959E-02 & 3.96012E+00 \\
 & & -7.24695E+00 & 3.70496E+00 & 2.81512E+00 \\
 $C_2^+$(total) & 1.84400E+01 \cite{locht} & 2.04011E-03 & -2.04011E-03 & 2.14350E-03 \\
 & & 1.62930E+00 & -2.41224E+00 & 1.76375E+00 \\
 $CH_2^+ + C^+ + 3e$ & 2.80000E+01 \cite{feil1}  & 1.10130E-09 & -1.10130E-09 & 4.87997E-01 \\
 & & -1.45604E+00 & 1.84000E+00 & -6.60842E-01 \\
 $CH^+$(total) & 2.08500E+01 \cite{davister1} & 9.40467E-13 & -9.36864E-13 & 4.50043E-10 \\
 & & 1.05143E+00 & -4.68811E-01 & 7.14727E-01 \\
 $C^+$(total) & 2.12000E+01 \cite{locht} & 2.90265E-18 & -2.90265E-18 & 6.15814E-02 \\
 & & -2.10614E-01 & 1.83046E+00 & -1.23394E+00 \\
 $C_2H_2^{2+} + 3e$ & 3.62000E+01 \cite{feil1} & 1.66952E-01 & -1.37602E-01 & -2.43008E-02 \\
 & & -1.39968E+00 & 9.24969E+00 & -8.49667E+00 \\
\hline
\end{tabular}
\end{table}

The results of the experiments on acetylene contain new data of one process which has not been included into 
HYDKIN so far, which is process \ref{c2h26}. However, the magnitude of that cross 
section is small ($\sim 2.7 \cdot 10^{-18} cm^2$ at maximum). The other new cross section fits have been 
compared to those available in the HYDKIN database by calculating the normalized root mean square deviation, d, 
for the entire ($E_{th}<E<1000eV$), low ($E_{th}<E<100eV$) and high energy range ($100eV<E<1000eV$). In 
addition, the deviation of the maximum values of the cross section fits relative to the corresponding HYDKIN 
curves, $\Delta \sigma_{max} = (\sigma_{new}^{max}-\sigma_{hyd}^{max})/\sigma_{hyd}^{max}$, as well as the 
energy shift of the maxima, $\Delta E_{max}=E_{max,new}-E_{max,hyd}$, i.e. the difference of the energy 
locations of the maxima of two corresponding cross section fits, have been calculated. The values of the
mean deviation as well as the deviation of the maxima and the maximum shift for each pair of corresponding 
curves are given in table \ref{table4}.

\begin{table}[h]
\renewcommand{\baselinestretch}{1}\normalsize
\caption{\label{table4}Accordance of present cross section fits with those based on earlier data compiled 
            by HYDKIN for the case of acetylene.  
	    Dissociation channels are indicated by product ions.}
\begin{tabular}{@{}lcccccc@{}}
\hline
 $C_2H_2 \rightarrow$ & $C_2H_2^+$ & $C_2H^+$ & $C_2^+$ & $CH^+$ & $C^+$ & $C_2H_2^{2+}$ \\
\hline
 $d$ in $\%$ for $E_{th} < E < 1keV$ & 3.3 & 11.0 & 13.4 & 19.9 & 49.5 & 101 \\
 $d$ in $\%$ for $E_{th} < E < 100eV$  & 6.3 & 30.0 & 13.1 & 36.3 & 27.8 & 270 \\
 $d$ in $\%$ for $100eV < E < 1keV$  & 3.0 & 7.6  & 13.4 & 18.0 & 50.9 & 80 \\
 $\Delta \sigma_{max}$ in $\%$       & -3.6 & 22.4 & 3.6 & -33.2 & -33.8 & 206  \\
 $\Delta E_{max}$ in $eV$            & 14.5 & -0.8 & 5.0 & 12.3  & -17.8 & -48.2 \\
\hline
\end{tabular}
\end{table}

\begin{figure}[h]
\includegraphics[width=.65\textwidth,height=75mm]{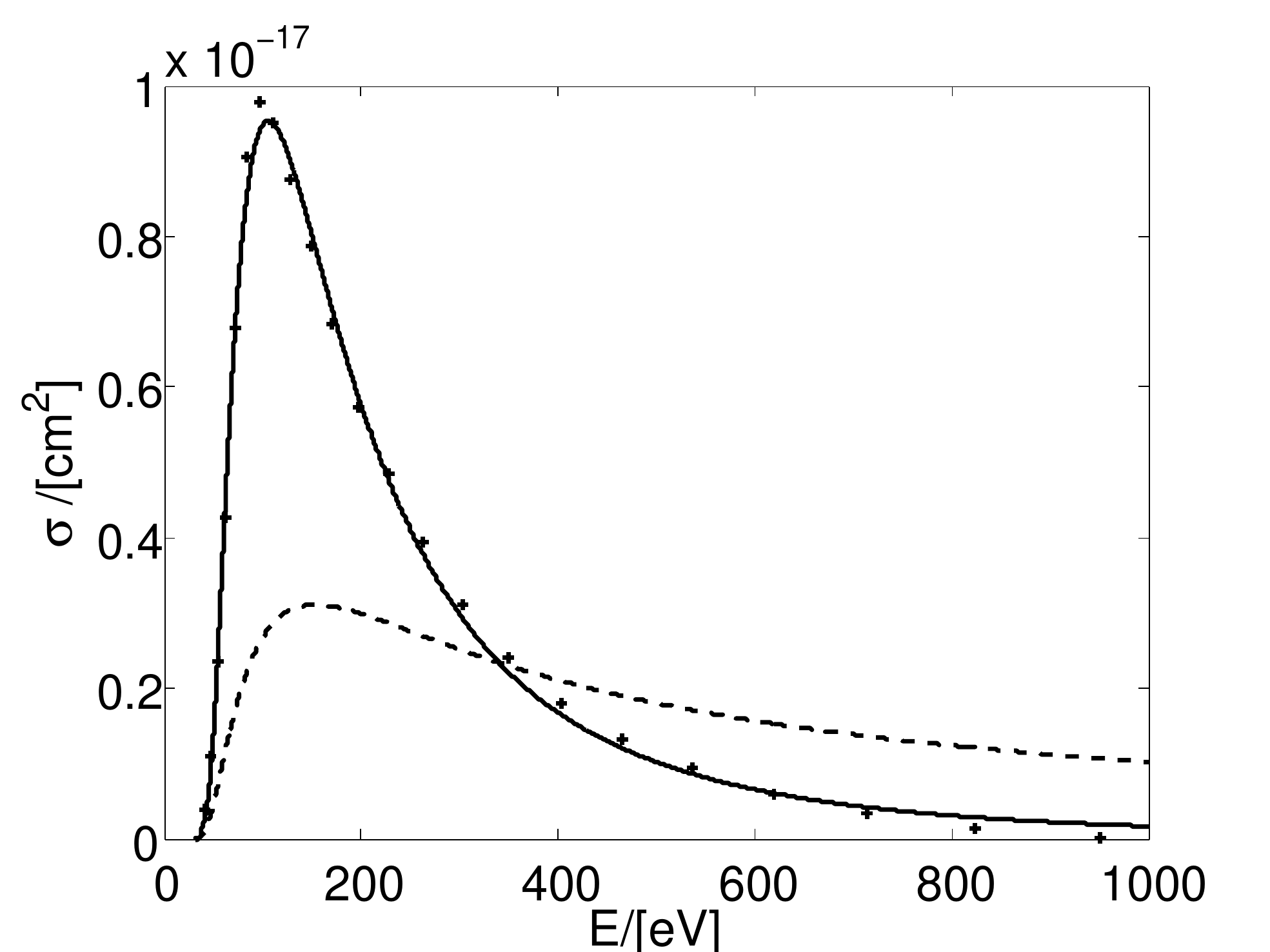}
\caption{\label{c2h2_c2h2++}Graphical comparison between new (solid line) and 
	  HYDKIN (dashed line) cross section fit for the process $e + C_2H_2 \rightarrow C_2H_2^{2+} + 3e$. The experimental data 
	  from Ref. \cite{feil1} is shown too (crosses).}
\end{figure}

The accordance of the curves is rather good, i.e. the deviation is less than the estimated experimental error 
of about $15\%$, for processes (\ref{c2h21},\ref{c2h23}) and the high energy range of process (\ref{c2h22}). 
For the rest of new cross section fits the deviations lie in a range of $18-51\%$, with exception of the fit 
for process (\ref{c2h27}) which shows a completely different energy dependence than the corresponding one 
provided by HYDKIN. \\
Also the direction of the deviation, i.e. if the new cross section fit is larger or smaller in magnitude 
than the HYDKIN one, is not unambigous. In case of process (\ref{c2h22}) the cross section is larger than 
the one provided by HYDKIN, where in case of processes (\ref{c2h25},\ref{c2h26}) it is smaller. \\
Most interesting is the energy behaviour of the cross section for process (\ref{c2h27}) which differs in many 
aspects from the corresponding one provided by HYDKIN. The new data fit for this process and the result from 
HYDKIN is shown in Fig. \ref{c2h2_c2h2++}. Comparing the magnitude 
of both cross section data fits a difference of $200\%$ can be observed near the maxima. Moreover, the 
maximum of the new data fit is located at about $50eV$ lower energy, leading thereby to a much steeper 
increase of the cross section right after the energy threshold. The most unexpected feature of process 
(\ref{c2h27}) is its steep decrease after the maximum, affecting also the assumption of approximately 
constant branching ratios, which will be discussed in more detail in Sec. \ref{invariance}. However, it might 
be interesting to note that the new cross section fit for the sum of processes (\ref{c2h25},\ref{c2h27}) 
accords to that one given by HYDKIN rather well within $20\%$, indicating thereby that the measurements of 
Feil et al. are in accordance with earlier cross section experiments for the $m/z=13$ Thompson dissociation 
channel, but the contribution of the two underlying processes determined by using cross sections differential 
to ion kinetic energy is different than expected.

\subsection{Ethylene}
\label{ethylene}
Ionization cross section data fits for the following electron-impact processes of ethylene, obtained by the 
measurements of Endstrasser et al. \cite{endstrasser}, have been generated by making use of Eq. 
\ref{fitformula}:
\begin{subequations}
 \begin{eqnarray}
  e + C_2H_4 & \rightarrow & C_2H_4^+ + 2e, \label{c2h41} \\
   & \rightarrow & C_2H_3^+ + H + 2e, \label{c2h42} \\
   & \rightarrow & C_2H_2^+ + \dots, \label{c2h43} \\
   & \rightarrow & C_2H^+ + \dots, \label{c2h44} \\
   & \rightarrow & C_2^+ + \dots, \label{c2h45} \\
   & \rightarrow & CH_3^+ + \dots, \label{c2h46} \\
   & \rightarrow & CH_2^+ + F + 2e. \label{c2h47} \\
   & \rightarrow & CH_2^+ + F^+ + 3e. \label{c2h48} \\
   & \rightarrow & CH^+ + \dots, \label{c2h49} \\
   & \rightarrow & C^+ + \dots, \label{c2h410} \\
   & \rightarrow & H_2^+ + \dots, \label{c2h411} \\
   & \rightarrow & H^+ + \dots, \label{c2h412} \\
   & \rightarrow & C_2H_3^{2+} + H + 3e, \label{c2h413} \\
   & \rightarrow & C_2H^{2+} + \dots, \label{c2h414}
 \end{eqnarray}
\end{subequations}
where $F$ denotes some unspecified neutral fragments and $F^+$ some unspecified fragment ions, and the dots 
again designate the sum of all possible dissociation sub channels. The fitting coefficients (together with 
threshold energies) for the cross section fits according to the processes listed above are given in table 
\ref{c2h4table}. The threshold energies for processes (\ref{c2h413},\ref{c2h414}) have been estimated during 
the fitting procedure, because no appropriate experimental values have been found.

\begin{table}[h]
\renewcommand{\baselinestretch}{1}\normalsize
\caption{\label{c2h4table}Values of fitting parameters $E_{th}$ (with references in squared brackets if 
            data have been appropriate and available) and $A_i$ in Eq. (\ref{fitformula}) for partial 
            ionization cross sections of ethylene. The cross sections marked with an asterisk have to be 
            corrected for a contribution of about $40\%$ of the $F$ cross section to the $F^+$ cross section
            \cite{huber}.}
\begin{tabular}{@{}lcccc@{}}
\hline                           
 Reaction $e + C_2H_4 \rightarrow$ & $E_{th} \ \textnormal{in} \ eV$ & $A_i, i=1-3$ &  \\
	 &	         & $A_i, i=4-6$ &  \\
\hline
 $C_2H_4^+ + 2e$ & 1.05100E+01 \cite{reiterkueppers,nist} & 1.55251E+00 & -1.42571E+00 & 3.33972E-01 \\
				   &	   & 1.92836E-01 & -3.85851E+00 & 2.77265E+00 \\
 $C_2H_3^+ + H + 2e$ & 1.30900E+01 \cite{reiterkueppers,nist} & 2.20951E+00 & -2.13713E+00 & -3.31816E-01 \\
 & & 8.37883E-02 & -7.02670E-01 & -2.15846E-01 \\
 $C_2H_2^+$(total) & 1.32300E+01 \cite{reiterkueppers,nist} & 1.42515E+00 & -1.40155E+00 & -3.05908E-02 \\
 & & 1.19451E+00 & -3.93545E+00 & 3.05752E+00 \\
 $C_2H^+$(total) & 1.90600E+01 \cite{reiterkueppers}  & 2.73180E-01 & -2.73180E-01 & -9.10600E-02 \\
 & & 1.46495E-01 & 2.95188E+00 & -2.77248E+00 \\
 $C_2^+$(total) & 2.45000E+01 \cite{nist} & 2.64453E-10 & -2.64453E-10 & 1.20273E-01 \\
 & & 3.35800E-01 & 5.22613E-01 & -5.03272E-01 \\
 $CH_3^+$(total) & 1.73400E+01 \cite{endstrasser} & 1.56138E-02 & -1.39287E-02 & -3.51951E-03 \\
 & & 9.73948E-02 & -1.94039E-01 & 1.07633E-01 \\
 $CH_2^+ + F + 2e$(total)$^*$
 & 1.88400E+01 \cite{endstrasser} & 1.05603E-01 & -7.50799E-02 & -4.67817E-03 \\
 & & 6.93006E-01 & -2.02644E+00 & 1.27088E+00 \\
 $CH_2^+ + F^+ + 3e$(total)$^*$ & 1.88400E+01 \cite{endstrasser} & 1.10167E-01 & -9.33446E-02 & -1.98998E-02 \\
 & & -2.47066E-01 & 1.67241E+00 & -8.94238E-01 \\
 $CH^+$(total) & 2.38700E+01 \cite{plessis} & 5.89932E-10 & -5.89932E-10 & 1.73032E-01 \\
 & & -1.67939E-01 & 1.74778E+00 & -1.22661E+00 \\
 $C^+$(total) & 2.70000E+01 \cite{plessis} & 1.33872E-09 & 6.79653E-03 & 4.20114E-01 \\
 & & -1.21662E+00 & 3.42678E+00 & -2.06732E+00 \\
 $H_2^+$(total), $N=8$ & 1.73700E+01 \cite{endstrasser} & 1.33848E-09 & 1.51608E-02 & 1.51608E-02 \\
 & & -6.06545E-01 & 3.35538E+00 & -8.14998E+00 \\
 & & 9.59769E+00 & -4.14264E+00 & \\
 $H^+$(total) & 1.88800E+01 \cite{endstrasser} & 8.98889E-03 & -8.98889E-03 & 3.33852E-01 \\
 & & -1.67362E+00 & 3.23495E+00 & -1.56285E+00 \\
 $C_2H_3^{2+} + H + 3e$ & 3.50000E+01 & 9.16389E-10 & 9.25783E-03 & 1.13076E-01 \\
 & & 4.85152E-01 & -8.08171E-01 & 3.48044E-01 \\
 $C_2H^{2+}$(total) & 5.00000E+01 & 2.71243E-12 & -1.86236E-12 & 8.41921E-03 \\
 & & -3.70929E-03 & -7.72609E-03 & 5.14723E-03 \\
\hline
\end{tabular}
\end{table}

\begin{figure}[h]
\includegraphics[width=.65\textwidth,height=75mm]{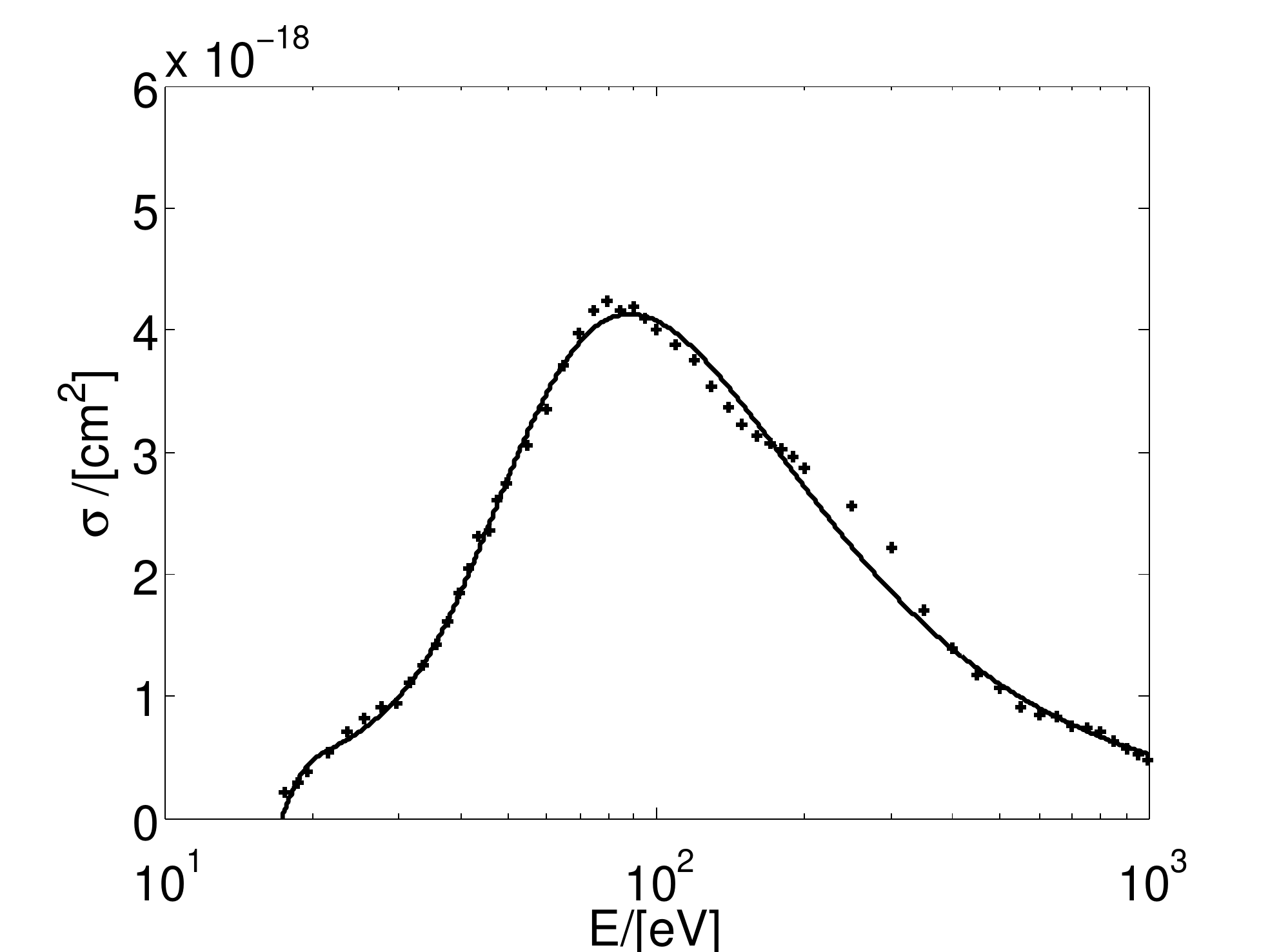}
\caption{\label{c2h4_h2+}Cross section fit (line) for the process $e + C_2H_4 \rightarrow H_2^+ + \dots$. 
         The experimental data \cite{endstrasser} is shown too (crosses).}
\end{figure}

The cross sections for processes (\ref{c2h47}-\ref{c2h49}) have been determined after subtracting the 
contributions from the ionization channels containing $C_2H_4^{2+},^{13}CCH_3^{2+}$ with $m/z=14$ for the 
$CH_2^+$ product ion channels and the contributions from the  $^{13}C^+,^{13}CCH^{2+},C_2H_2^{2+}$ channels 
with $m/z=13$ for the $CH^+$ product ion channel, since it is only possible to 
distinguish product ions with respect to their mass to charge ratio with a mass spectrometer. In the former 
case, the contribution has been estimated to be about $12\%$ of the total $m/z=14$ product ion channel, and 
in the latter case to be about $11\%$ of the total $m/z=13$ product ion channel, see also Ref. 
\cite{endstrasser}. A distinction between processes (\ref{c2h47},\ref{c2h48}) which contain the same fragment 
ion $CH_2^+$ has been possible by making use of the corresponding cross sections differential to initial ion 
kinetic energy given in Ref. \cite{endstrasser}. Therein the quasithermal part ($<0.5eV$) of the cross section 
has been assigned to direct dissociative ionization, i.e. process (\ref{c2h47}), and the high 
energy and dominating part of the cross section to process (\ref{c2h48}). The latter is expected to be most 
likely produced via Coulomb explosion of the doubly charged parent ion \cite{endstrasser}:
\begin{equation}
 e + C_2H_4 \rightarrow C_2H_4^{2+} + 3e \rightarrow CH_2^+ + F^+ +3e,
\end{equation}
where $F^+$ denotes a set of fragments with one of them being ionized, e.g. $F^+ = C^+ + H_2$. \\
For process (\ref{c2h411}) it has been necessary to increase the number of fitting coefficients to $N=8$ to 
obtain a physically well behaving fit. However, the near threshold behaviour of this cross section fit is 
untypical as well as the cross section fit of process (\ref{c2h412}), see figure \ref{c2h4_h2+}. This might 
be an indication that these cross sections are superpositions of two distinct 
cross sections for different processes with completely different threshold energies. Possible processes are 
direct dissociative ionization with fragment ion $H_2^+$ or $H^+$, respectively, and some Coulomb explosion 
of an intermediate doubly charged ion with a much enhanced energy threshold in the range of $30-50eV$.

As for the previously discussed molecule acetylene, also for ethylene the cross section fits for 
(dissociative) ionization have been compared to earlier data compiled by HYDKIN. The values for 
$d, \Delta \sigma_{max}$ and $\Delta E_{max}$ are given in table \ref{table5}.

\begin{table}[h]
\renewcommand{\baselinestretch}{1}\normalsize
\caption{\label{table5}Accordance of present cross section fits with those provided by HYDKIN for the case 
            of ethylene. Dissociation channels are indicated by product ions (i.e. no distinction between 
            different dissociation subchannels wrt to neutral fragmentation; if the HYDKIN cross section 
            belongs to a specific dissociation subchannel, the neutral fragments are given in parenthesis).}
\begin{tabular}{@{}lccccc@{}}
\hline
 $C_2H_4 \rightarrow$ & $C_2H_4^+$ & $C_2H_3^+$ & $C_2H_2^+$ & $C_2H^+(H_2,H)$ & $C_2^+(2H_2)$ \\
\hline
 $d$ in $\%$ for $E_{th} < E < 1keV$    & 38.3   & 14.7 & 18.0  & 43.2   & 56.8 \\
 $d$ in $\%$ for $E_{th} < E < 100eV$  & 27.4   & 42.4 & 47.3  & 100.6  & 90.5 \\
 $d$ in $\%$ for $100eV < E < 1keV$     & 39.2   & 8.7  & 12.4  & 34.3   & 53.3 \\
 $\Delta \sigma_{max}$ in $\%$          & -29.0  & -3.8 & -13.0 & -2.3   & 87.3  \\
 $\Delta E_{max}$ in $eV$               & -26.0  & -9.6 & -2.2  & -44.2  & -11.7     \\
\hline
 $C_2H_4 \rightarrow$ & $CH_3^+(CH)$ & $CH_2^+(CH_2)$ & $CH^+(CH_3)$ & $C^+$ \\
\hline
 $d$ in $\%$ for $E_{th} < E < 1keV$ & 88.3  & 51.4   & 138.4 & 286.7 \\
 $d$ in $\%$ for $E_{th} < E < 100eV$  & 85.8  & 61.3   & 183.5 & 313.3 \\
 $d$ in $\%$ for $100eV < E < 1keV$  & 88.5  & 50.6   & 134.3 & 284.5 \\
 $\Delta \sigma_{max}$ in $\%$       & -89.1 & -47.7  & 191.1 & 358.0  \\
 $\Delta E_{max}$ in $eV$            & -38.9 & -52.4  & -9.6  & -2.5 \\
\hline
\end{tabular}
\end{table}

The comparison of the remaining set of new cross section data fits for ethylene with counterparts compiled by 
HYDKIN has shown much more significant deviations as for the case acetylene. Only the cross section fits for 
processes (\ref{c2h42},\ref{c2h43}) accord to their counterparts 
provided by HYDKIN in the high energy range $100eV < E < 1keV$ within the estimated experimental error of 
about $15\%$. The deviations at low energies range from $\sim 30\%$ up to $\sim 350\%$, with exception of the 
two above mentioned processes not quite different from the range of the deviations at high energies. One might 
argue that the cross section fits provided by HYDKIN are only given for distinct dissociation channels and a 
comparison to total ones (with respect to neutral fragmentation) neglects the contribution of other possible 
channels, but since most of the present (total) cross sections are actually smaller than the (partial) ones 
from HYDKIN this cannot be a sufficient explanation for all the deviations. However, most of the differences 
are still within $100\%$, with exception of processes (\ref{c2h49},\ref{c2h410}), which are 
about a factor of 3 and 4, respectively, larger compared to the HYDKIN curves. The magnitude of these 
processes is larger than their counterparts in HYDKIN. This is also the case for process (\ref{c2h45}), which 
is about a factor of $\sim 1.5-2$ larger than its counterpart in HYDKIN. Among these three processes no other 
process with that property has been found. Possible reasons for the large deviations in 
contrast to the former case of acetylene could be due to the fact that the cross sections provided by HYDKIN 
for ethylene are based on experimental data only at two energies (75eV and 3.5MeV) \cite{jrethanpropanpaper}. \\
It might also be interesting to note that the present cross sections maxima are all shifted to lower energies 
compared to HYDKIN, indicating thereby a steeper increase right beyond the energy threshold. These shifts 
range from very small values of about 2eV up to $\sim$50eV. \\
Since cross section fits for processes (\ref{c2h48},\ref{c2h411}-\ref{c2h414}) have so far not been available 
in HYDKIN they do not occur in table \ref{table5}. It is noted here that these 
cross sections are rather small compared to the dominating ones, as is the case for process (\ref{c2h24}) 
occuring in the catabolism of acetylene.

\subsection{Propene}
\label{propene}
Ionization cross section data fits for the following electron-impact processes of propene, obtained by the 
measurements of Feil et al. \cite{feil2}, have been generated:
\begin{subequations}
 \begin{eqnarray}
  e + C_3H_6 & \rightarrow & C_3H_6^+ + 2e, \label{c3h61} \\
   & \rightarrow & C_3H_5^+ + H + 2e, \label{c3h62} \\
   & \rightarrow & C_3H_4^+ + \dots, \label{c3h63} \\
   & \rightarrow & C_3H_3^+ + \dots, \label{c3h64} \\
   & \rightarrow & C_2H_3^+ + F + 2e, \label{c3h65} \\
   & \rightarrow & C_2H_3^+ + F^+ + 3e, \label{c3h66} \\
   & \rightarrow & C_2H_2^+ + F^+ + 3e, \label{c3h67} \\
   & \rightarrow & C_2H_2^+ + F'^+ + 3e, \label{c3h68} \\
   & \rightarrow & C_2H^+ + F^+ + 3e, \label{c3h69} \\
   & \rightarrow & C_2H^+ + F'^+ + 3e, \label{c3h610} \\
   & \rightarrow & CH_3^+ + F + 2e, \label{c3h611} \\
   & \rightarrow & CH_3^+ + F^+ + 3e, \label{c3h612} \\
   & \rightarrow & CH_2^+ + F + 2e, \label{c3h613} \\
   & \rightarrow & CH_2^+ + F^+ + 3e, \label{c3h614} \\
   & \rightarrow & CH^+ + F^+ + 3e, \label{c3h615} \\
   & \rightarrow & CH^+ + F'^+ + 3e, \label{c3h616} \\
   & \rightarrow & C^+ + \dots, \label{c3h617}
 \end{eqnarray}
\end{subequations}
where $F$ again denotes some unspecified fragments, $F^+$ some unspecified fragment ions, $F'^+$ some highly 
excited unspecified fragment ions, and the dots have the same interpretation as in the previous sections. The 
final fitting coefficients are summarized in table \ref{c3h6table}.

\begin{table}[h]
\renewcommand{\baselinestretch}{1}\normalsize
\caption{\label{c3h6table}Values of fitting parameters $E_{th}$ (with references in squared brackets; if 
            data have been appropriate and available) and $A_i$ in Eq. (\ref{fitformula}) for partial 
            ionization cross sections of propene.}
\begin{tabular}{@{}lcccc@{}}
 \hline
 Reaction $e + C_3H_6 \rightarrow$ & $E_{th} \ \textnormal{in} \ eV$ & $A_i, i=1-3$ &  \\
	 &	         & $A_i, i=4-6$ &  \\
\hline
 $C_3H_6^+ + 2e$ & 9.73000E+00 \cite{reiterkueppers,nist} & 1.76465E+00 & -1.76444E+00 & -7.19486E-01 \\
				   &	   & 1.06644E+00 & 4.59632E-02 & -1.79033E+00 \\
 $C_3H_5^+ + H + 2e$ & 1.19000E+01 \cite{nist} & 4.78140E+00 & -4.78140E+00 & -2.27612E+00 \\
 & & -2.81997E-01 & 6.18934E+00 & -8.94871E+00 \\
 $C_3H_4^+$(total) & 1.19100E+01 \cite{nist} & 1.17937E+00 & -1.17937E+00 & -3.93124E-01 \\
 & & -5.00436E-01 & 1.82754E+00 & -2.00461E+00 \\
 $C_3H_3^+$(total) & 1.31900E+01 \cite{nist}  & 2.38603E+00 & -2.38311E+00 & -8.85054E-02 \\
 & & -4.55078E+00 & 1.03029E+01 & -6.87145E+00 \\
 $C_2H_3^+ + F + 2e$(total) & 1.37000E+01 \cite{nist} & 2.08093E-01 & -2.08093E-01 & 3.85608E-02 \\
 & & -1.36777E+00 & 3.77297E+00 & -2.65960E+00 \\
 $C_2H_3^+ + F^+ + 3e$(total) & 3.20000E+01 & 1.02861E+00 & -9.39380E-01 & -2.53637E-01 \\
 & & 9.24639E-01 & 1.93228E+00 & -3.84679E+00 \\
 $C_2H_2^+ + F^+ + 3e$(total) & 2.70000E+01 & 3.87573E-02 & -1.28254E-02 & 1.30128E-02 \\
 & & 7.99346E-02 & 2.53853E-01 & -3.39084E-01 \\
 $C_2H_2^+ + F'^+ + 3e$(total) & 3.80000E+01 & 4.18718E-01 & -3.74737E-01 & -9.55924E-02 \\
 & & -2.30625E-01 & 1.90802E+00 & -2.32656E+00 \\
 $C_2H^+ + F^+ + 3e$(total) & 3.45000E+01 & 3.70574E-10 & -3.66193E-10 & 1.72848E-01 \\
 & & -5.42785E-01 & 1.03012E+00 & -6.24479E-01 \\
 $C_2H^+ + F'^+ + 3e$(total) & 5.00000E+01 & 1.38462E-01 & -1.27538E-01 & -1.96257E-02 \\
 & & 8.06484E-02 & 6.26341E-01 & -9.10837E-01 \\
 $CH_3^+ + F +2e$(total) & 1.70000E+01 & 2.92264E-11 & 5.00748E-04 & 3.43137E-02 \\
 & & 1.08047E+00 & -2.72127E+00 & 1.72080E+00 \\
 $CH_3^+ + F^+ + 3e$(total) & 2.50000E+01 & 5.10937E-01 & -5.10937E-01 & -1.70312E-01 \\
 & & 1.08936E+00 & -7.24946E-01 & -6.60811E-01 \\
 $CH_2^+ + F + 2e$(total) & 1.70000E+01 \cite{nist} & 4.61937E-02 & -4.61937E-02 & -1.53979E-02 \\
 & & 1.04334E-01 & -7.55917E-02 & -9.05605E-02 \\
 $CH_2^+ + F^+ + 3e$(total) & 2.50000E+01 & 4.01244E-01 & -4.01244E-01 & 3.00150E-01 \\
 & & -2.88555E+00 & 6.71214E+00 & -4.69804E+00 \\
 $CH^+ + F^+ + 3e$(total) & 2.40000E+01 & 1.60325E-02 & -1.60325E-02 & -5.34418E-03 \\
 & & 9.37462E-02 & -1.31281E-01 & 3.89138E-02 \\
 $CH^+ + F'^+ + 3e$(total) & 2.75000E+01 & 4.73250E-02 & -4.73250E-02 & 5.46693E-02 \\
 & & -6.65212E-01 & 2.21333E+00 & -1.57914E+00 \\
 $C^+$(total) & 3.00000E+01 & 1.66755E-10 & -1.66755E-10 & 3.29712E-01 \\
 & & -1.68434E+00 & 3.39804E+00 & -1.90085E+00 \\
 \hline
\end{tabular}
\end{table}

Cross section fits for the processes (\ref{c3h65}-\ref{c3h616}) have been obtained by analyzing the numerous 
cross sections differential to initial ion kinetic energy measured in Ref. \cite{feil2}. Within this family 
of cross sections there exist many where the threshold energies appear to be too large to assign it to direct 
dissociative ionization. For this reason, Coulomb explosion of doubly charged intermediate ions might be a 
more reasonable assumption for explaining the high threshold energies. Since the measured cross sections are 
total ones with respect to neutral fragmentation channels, a further specification of $F$ and $F'$ would be 
highly speculative at this stage. However, it has been possible to distinguish between direct dissociative 
ionization and Coulomb explosion channels in the cases of processes (\ref{c3h65},\ref{c3h611}-\ref{c3h614}). 
For processes (\ref{c3h67}-\ref{c3h610},\ref{c3h615},\ref{c3h616}) the threshold energies of both 
dissociation channels are in range of Coulomb explosion channels, but the appropriate differential cross 
sections have shown a contribution of distinct fragmentation channels, which is indicated by the distinction 
between $F$ and $F'$. With exception of process (\ref{c3h613}) the threshold energies for processes 
(\ref{c3h66}-\ref{c3h617}) have been estimated during the fitting procedure, because no appropriate 
experimental values have been found. The numerical values for all threshold energies are given in Table 
\ref{c3h6table}.

\begin{figure}[h]
\includegraphics[width=.65\textwidth,height=75mm]{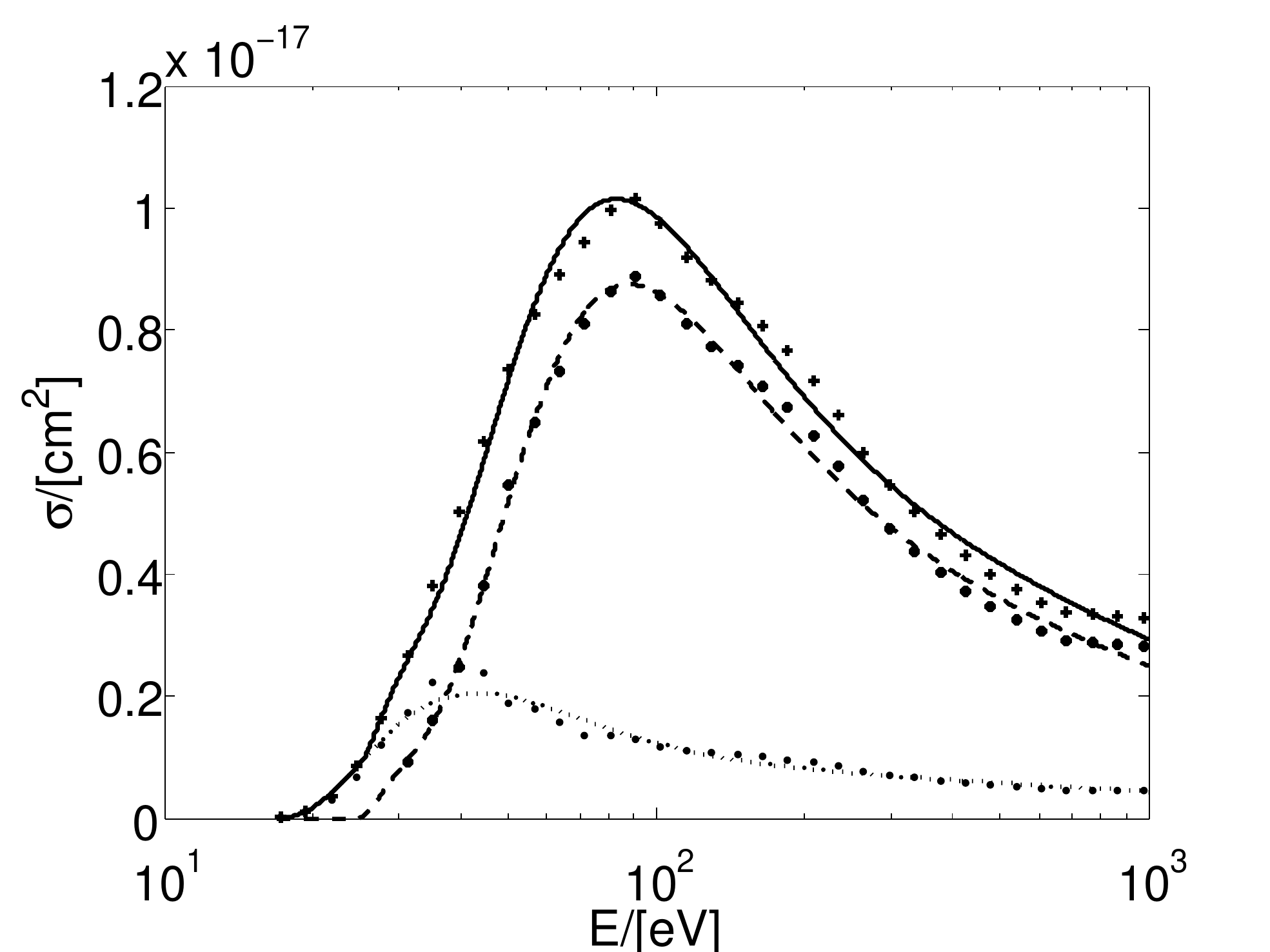}
\caption{\label{ce}Cross sections differential to ion kinetic energy: Experimental data 
	 and fits. Total cross section for the process 
	 $e + C_3H_6 \rightarrow CH_2^+ + \dots$ (crosses and solid line), contribution from direct 
	 dissociative ionization (dots and dotted line), contribution from Coulomb explosion (asterisks and dashed line).}
\end{figure}

In figure \ref{ce} an example for the significant role of Coulomb explosion is shown: In case of ionization 
of $C_3H_6$ leading to the product ion $CH_2^+$ the Coulomb explosion channel is actually the dominating one, 
i.e. at higher energies (at least greater than $\sim 50eV$) the probability that $CH_2$ originates from 
electron-impact ionization via Coulomb explosion is much enhanced relative to direct dissociative ionization 
with $CH_2^+$ being the only charged fragment. \\
An analogous comparison as in the previous sections has been done for seven ionization channels of propene, 
because all the other ionization channels (in total 10) were not included in the earlier data compiled by 
HYDKIN. The values for the normalized root mean square deviations $d$ as well as the deviation of the maxima 
and the maximum shifts of the appropriate cross sections can be found in table \ref{table7}.

\begin{table}[h]
\renewcommand{\baselinestretch}{1}\normalsize
\caption{\label{table7}Accordance of present cross section fits with those provided by HYDKIN for the case 
            of propene: The appropriate processes are indicated by product ions (i.e. no distinction between 
            different dissociation subchannels wrt to neutrals; if the HYDKIN cross section belongs to a 
            specific dissociation channel, the neutral fragments are given in parenthesis).}
\begin{tabular}{@{}lcccc@{}}
\hline
 $C_3H_6 \rightarrow$ & $C_3H_6^+$ & $C_3H_5^+$ & $C_3H_4^+(H_2)$ & $C_3H_3^+(H_2,H)$ \\
\hline
 $d$ in $\%$ for $E_{th} < E < 1keV$ & 32.5 & 214.6  & 10.2  & 121.2 \\
 $d$ in $\%$ for $20eV < E < 100eV$  & 20.9 & 151.4  & 20.9  & 107.8 \\
 $d$ in $\%$ for $100eV < E < 1keV$  & 33.4 & 219.4  & 8.6   & 122.3 \\
 $\Delta \sigma_{max}$ in $\%$       & 16.0 & 138.1  & 23.6  & 103.5  \\
 $\Delta E_{max}$ in $eV$            & -9.4 & -5.9   & -2.8  & -3.6     \\
\hline
 $C_3H_6 \rightarrow$ & $C_2H_3^+(CH_3)$ & $CH_3^+(C_2H_3)$ & $CH_2^+(C_2H_4)$ & \\
\hline
 $d$ in $\%$ for $E_{th} < E < 1keV$    & 78.4  & 91.8  & 78.9 & \\
 $d$ in $\%$ for $20-30eV < E < 100eV$  & 74.0  & 87.6  & 70.1 & \\
 $d$ in $\%$ for $100eV < E < 1keV$     & 78.8  & 91.1  & 79.5 & \\
 $\Delta \sigma_{max}$ in $\%$          & -71.7 & -82.1 & -67.7 &  \\
 $\Delta E_{max}$ in $eV$               & -14.3 & -45.2 & -35.8 &     \\
\hline
\end{tabular}
\end{table}
 
With the single exception of the high energy range of process (\ref{c3h63}), where the 
accordance between the new cross section fits and earlier data fits compiled by HYDKIN is rather good 
($8.6\%$), the deviations between HYDKIN curves and 
current results are out of the range of the estimated experimental error of $15\%$. The closest matching has 
been found for process (\ref{c3h61}), where the deviation is $\sim30\%$. In all other cases the difference in 
magnitude is about a factor of 2-5. For fragment ions of the type $C_3H_y$ the recent 
measurements show larger values than the HYDKIN curves, where for the dissociation channels with fragment 
ions $C_2H_3^+, CH_3^+$ and $CH_2^+$ the magnitude of the cross section is significantly smaller. As it has 
already been mentioned in the case of acetylene, these deviations would be smaller if no distinction between 
direct dissociative ionization and fragmentation due to Coulomb explosion was possible. Then, the sum of 
these different processes would be assigned to direct dissociative ionization. One can therefore conclude 
that the process of Coulomb explosion has a more important role than expected, especially for higher 
hydrocarbons. This statement is also confirmed by the numerous processes listed in the previous 
sections that have been allocated to such processes due to their high energy thresholds. However, this effect 
of doubly charged intermediate ions makes the situation more complex concerning the assignment of cross 
sections to distinct fragmentation channels. \\
As a secondary remark it is noted that as in the case of ethylene, it has been observed that all of the cross 
sections maxima are shifted to lower energies compared to HYDKIN cross sections, and indicate therefore again 
a steeper increase right beyond the threshold.

\subsection{Branching ratios}
\label{invariance}
An ingredient for the assembly of the cross section database HYDKIN has been the assumption of 
(approximately) energy invariant branching ratios for multichannel processes in the energy region above 
$20-30eV$ \cite{jrethanpropanpaper}.For (dissociative) ionization processes going through a dipole allowed 
transition the energy invariance is based on the Born-Bethe behaviour of the cross section for large 
energies. Only in those cases when (dissociative) ionization proceeds by a different (non-dipole allowed) 
transition, e.g. a Coulomb explosion, then the typical $ln(cE)/E$ behaviour at high energies cannot be taken. 
An investigation of the energy dependence of the branching ratios based on the recent experimental data 
could thus illuminate the underlying dissociation mechanisms.
The branching ratios $R_j(E)$ are defined, see Ref. \cite{jrethanpropanpaper}, to
\begin{equation}
 R_{C_{x'}H_{y'}^q}(E)=\frac{\sigma_{C_{x'}H_{y'}^+}^{part}(E)}{\sigma_{C_xH_y}^{tot}(E)},
\end{equation}
where the j-th ionization channel is indicated by the product ion $C_{x'}H_{y'}^q$, 
$\sigma_{C_xH_y}^{tot}(E)$ denotes the total ionization cross section for the parent molecule 
$C_xH_y$, $\sigma_{C_{x'}H_{y'}^q}^{part}(E)$ the partial ionization cross section according to the process 
$e+C_xH_y\rightarrow C_{x'}H_{y'}^q + \dots$, and $q$ denotes the charge state of the 
product ion, i.e. here $q=+,2+$. Obviously, $x' \leq x$ and $y' \leq y$. As the recent measurements 
have not been sensitive to distinguish fragmentation channels with respect to neutrals, the specification of 
the reaction channel with respect to product ions is sufficient. \\
For acetylene seven distinct reaction channels have been measured and data fits have been generated. 
Dividing these by the total ionization cross section, i.e. the sum of partial ones, seven branching ratios 
have been determined and compared to appropriate counterparts in HYDKIN. It is interesting to note that both, 
present branching ratios and those obtained from HYDKIN clearly depart from energy invariance. Considering 
for instance the $C_2^+,CH^+,C^+$ reaction channels the branching ratios 
appear to approach constant asymptotic values (if at all) only very slowly and far beyond $10keV$. In other 
cases, as the dominating $C_2H_2^+,C_2H^+$ channels, the assumption of energy invariant branching ratios 
appears quite more reasonable, although only within certain limits. The branching ratio for the $C_2H_2^{2+}$ 
yields, however, an absolutely different behaviour compared to the HYDKIN counterpart, increasing again 
actually beyond $\sim 2keV$. On balance, the typical energy dependence of branching ratios (with few 
exceptions as the parent ion channel) can be characterized by a sharp increase 
after the energy threshold, a broad maximum at $20-200eV$ and a slow decrease afterwards. \\
In order to derive quantitative criteria 
for an approximation of constant branching ratios to be reasonable, the maximum deviations, $\delta_{max}$, 
from the mean values of the branching ratios, $<R_j>$, have been calculated according to
\begin{subequations}
\begin{eqnarray}
 \delta_{max} & = & max\left(\left|1-\frac{<R_j>}{R_j(E)}\right|\right), \\ 
 <R_j> & = & \frac{1}{E_{max}-E_{min}}\int_{E_{min}}^{E_{max}}R_j(E)dE,
\end{eqnarray}
\end{subequations}
where $j$ indicates again the reaction channel, $E_{max}=10keV$, 
$E_{min}=E_{th,max}+\Delta E \in (40,60,50)eV$ for $(C_2H_2$, $C_2H_4$, $C_3H_6)$, respectively. The latter 
has been chosen such that all reaction channels are open and considerable beyond the largest energy 
threshold to ensure a negligible influence of the energy threshold on the energy dependence of the branching 
ratio. \\
The two dominating $C_2H_2^+$,$C_2H^+$ channels as well as the $CH_2^+$ channel 
have smaller $\delta_{max}$ ($<22\%$), where for the rest of the subdominant channels $\delta_{max}=30-50\%$, 
with exception of the $C_2H_2^{2+}$ channel with $\delta_{max}=315\%$. \\
For the 13 distinct reaction channels in case of ethylene an analogous 
analysis of their branching ratios has been made. 
The branching ratios have mostly the typical behaviour as described above, with 
exception of the parent ion and the $C_2H_3^+$,$C_2H_2^+$ reaction channels, where in the two latter cases the 
branching ratio decreases somewhat beyond maximum, but then increases again after reaching a local minimum at 
$\sim 100eV$. Considering the $H_2^+$,$H^+$ reaction channels, one can again observe the unusual behaviour 
right beyond the energy threshold, which might be an indication that these cross sections are superpositions 
of cross sections referring to very different energy thresholds, see Sec. \ref{ethylene} \\
Considering the maximum deviations from the mean, $\delta_{max}$, it has been observed that these are lower 
for the most dominant channels than for less dominant ones: The $C_2H_4^+$,$C_2H_2^+$ channels have values of 
$\delta_{max}$ even below $5\%$ (together with the $CH_3^+$ channel, however, being much smaller in magnitude), 
followed by the $CH_2^+$, $C_2H_3^+$ channels with $\delta_{max}=12.8$ and $\delta_{max}=20.8$, respectively. 
The rest of the maximum deviations are in a range of $37.9-66.8\%$. Separating reaction channels into dominant 
and subdominant ones, it appears that the assumption of energy 
invariant branching ratios would cause less failure (if the cross sections would be unknown and instead 
computed by applying scaling laws to known ones) in case of the dominant ones than in case of the subdominant 
ones. This is supported also by the analysis for acetylene, where the ionization is mostly 
dominated by only two reaction channels, namely the $C_2H_2^+,C_2H^+$ channels, which add up to more than 
$90\%$ of the total cross section at almost all energies.In view of underlying dissociation mechanisms, one 
might conclude that the dominating reaction channels are mainly given by dipole allowed transitions, whereas 
the stronger variation of the branching ratio with energy in case of subdominant reaction channels 
indicates greater involvement of alternative (non-dipole allowed) dissociation mechanisms.\\
Also interesting to note is the fact that the recent measurements for (dissociative) ionization of ethylene 
show a different ordering of contributing reaction channels with respect to the magnitude of the appropriate 
branching ratio compared to the ordering of cross sections given by HYDKIN. In the latter case the ordering 
is as follows, beginning with the largest contributing reaction channel: $C_2H_4^+$, $C_2H_3^+$, $C_2H_2^+$, 
$C_2H^+$, $CH_3^+$, $CH_2^+$, $C_2^+$, $CH^+$, $C^+$. Instead of this ordering the recent measurements yield the 
following: $C_2H_4^+$, $C_2H_3^+$, $C_2H_2^+$, $C_2H^+$, $CH_2^+$, $CH^+$, $C^+$, $C_2^+$, $H^+$, $H_2^+$, $C_2H_3^{2+}$,
 $CH_3^+$, $C_2H^{2+}$, where it has to be mentioned that the contribution of the $C_2H_3^+$ channel actually 
becomes larger than that of the $C_2H_4^+$ channel for energies beyond $1keV$ as well as the 
contribution of the $CH_2^+$ channel becomes larger than that of the $C_2H^+$ channel for energies beyond 
$200eV$. Therefore, the ordering of contributing reaction channels to the total ionization cross section in 
case of ethylene is different compared to that one based on earlier data compiled by HYDKIN with respect to 
the following aspects:
\begin{itemize}
 \item[(a)] According to the recent measurements there exist intersections between branching ratios for pairs 
            of processes $C_2H_4^+$, $C_2H_3^+$ and $C_2H^+$, $CH_2^+$ in contrast to the partial cross section 
            fits provided by HYDKIN, where the magnitude of the contribution of the $C_2H_3^+$ and $CH_2^+$ 
            channel to the total cross section is for all energies smaller than the contribution of the 
            $C_2H_4^+$ and $C_2H^+$ channel, respectively. In addition are the contributions 
            of the $C_2H_4^+$, $C_2H_3^+$ channels different by a factor of $1.5-2$ in the case of HYDKIN, 
            whereas they are approximately equal according to the recent measurements for energies larger 
            than $\sim 100eV$.
 \item[(b)] The $CH_3^+$ channel is on the fifth position in the HYDKIN ordering, whereas it appears next to 
            the last in the present ordering. A small contribution appears reasonable also because of the 
            molecular structure of ethylene ($H_2C$=$CH_2$), because production of a $CH_3^+$ ion gives rise 
            to a rearrangement reaction chain like 
            $e + H_2C$=$CH_2 \rightarrow e + H$:$C$-$CH_3^* \rightarrow CH_3^+ + \dots$.
 \item[(c)] The ordering of the subdominant $C_2^+$,$CH^+$,$C^+$ channels according to HYDKIN, is 
            $CH^+$, $C^+$,$C_2^+$ according to the recent experiments.
 \item[(d)] For completeness it is noted that the present ordering includes also the $H_2^+$,$H^+$, $C_2H_3^{2+}$, 
            $C_2H^{2+}$ reaction channels, which have so far not been available in the HYDKIN database. It is 
            also noted that the first three of them are larger than the $CH_3^+$ channel (see (b)).
\end{itemize}
However, the ordering of the three most dominating reaction channels has been untouched below $1keV$, although 
their contribution to the total cross section is partitioned in another way. Noting that they add 
up to almost $90\%$ of the total cross section, the changes of ordering and magnitude of the other channels 
are just small modifications on the scale of the total cross section. \\
An analogous analysis of the branching ratios for the 11 reaction channels, distinguished 
by product ions, measured for propene has been made. 
Beside the fact that the determined branching ratios for the $C_2H_3^+$, $C_2H_2^+$, $C_2H^+$, $CH^+$, $C^+$ reaction 
channels have larger deviations from the mean of branching ratios (in a range of $35-154\%$), the six 
remaining channels (containing the dominating ones) have only a rather small deviation, $\delta_{max}<23\%$. 
A smaller deviation from the mean of branching ratios has been found for the dominating reaction channels in 
cases of acetylene and ethylene and this holds also for the dominating reaction channels in the case of 
propene. \\
However, the magnitudes as well as the ordering of the branching ratios differ 
quite strongly from those obtained from HYDKIN. Whereas the four most dominating reaction channels add up to 
about $80-90\%$ of the total cross section, they do just for about half in HYDKIN. Many of the subdominant 
channels are therefore much smaller compared to their HYDKIN counterparts. The complete ordered 
set (beginning 
with the largest contribution) of contributing reaction channels given by HYDKIN \cite{reiterkueppers} reads: $C_3H_6^+$,
$C_3H_5^+$, $C_3H_4^+$, $C_3H_3^+$, 
$C_2H_3^+$, $C_2H_4^+$, $C_2H_2^+$, $CH_3^+$, $C_2H_5^+$, $C_3H_2^+$, $CH_4^+$, $CH_2^+$, $CH^+,$ where the 
$C_2H_4^+$, $C_2H_2^+$ and 
the $CH_3^+$, $C_2H_5^+$ reaction channels interchange their positions for energies below $30eV$ and $70eV$, 
respectively. In contrast, the complete ordered set (again beginning with the largest contribution) of 
contributing reaction channels obtained from the recent measurements reads: $C_3H_5^+$, $C_3H_6^+$, $C_3H_3^+$, 
$C_3H_4^+$, $C_2H_3^+$, $C_3H_2^+$, $C_3H^+$, $CH_3^+$, $CH_2^+$, 
$C_2H_2^+$, $C_2H_4^+$, $CH^+$, $C^+$, $C_2H^+$, where the $C_3H_6^+$, $C_3H_3^+$ channels contribute almost equally 
to the total cross section, the $C^+$, $C_2H^+$ channels interchange their position for energies above 
$\sim2.5keV$, and the $C_3H_2^+$, $C_3H^+$, $C_2H_4^+$ channels have been added due to the peaks at 
$m/z=38,37,28$ Thompson, respectively, in the mass spectrum recorded at an electron energy of $100eV$ by 
Feil et al \cite{feil2}. However, there have been reported no cross 
sections for these values of $m/z$ and their contribution has been estimated to be $\sim 10-20\%$ to the 
total cross section. \\
Comparison of these ordered sets and analysis of magnitudes of the branching ratios yields that the 
differences in magnitude and ordering of the contributing reaction channels to ionization of propene between 
present reaction channels and HYDKIN ones can be summarized in the following:
\begin{itemize}
 \item[(a)] The ordering of the $C_3H_y,y \in \{3,4,5,6\}$ (and most dominating) channels yields enhanced 
            contribution for odd $y$ compared to the HYDKIN ordering from $y=6$ counting straight down to 
            $y=3$. Indeed, the contribution of the $C_3H_5^+$ channel is the largest, quite two times that 
            of the $C_3H_6^+$, $C_3H_3^+$ channels (whose contribution is almost equal) and quite four times 
            that of the $C_3H_4^+$ channel. Together they add up to about $80-90\%$ of the 
            total ionization cross section of propene.
 \item[(b)] There exist $C_2H_5^+$, $CH_4^+$ channels in the ordered set of HYDKIN, which do not seem to have 
            any non-negligible contribution to the total cross section at all according to the recent 
            measurements.
 \item[(c)] Also the subdominant reaction channels differ as well in magnitude as in ordering.
\end{itemize}

\section{Conclusion}
\label{summary}
Partial electron impact ionization cross sections have been analyzed for acetylene, ethylene and propene. 
Data fits are presented in analytic forms according to Eq. (\ref{fitformula}) and have been compared to 
earlier data compiled by the online cross section database HYDKIN. \\ 
For acetylene, rather good accordance is found in view of the dominating reaction channels, and larger 
deviations in case of sub dominating channels. One new cross section fit, i.e. one which has not been included 
in the database so far, for the $(CH_2^+ + C^+)$ channel has been generated. \\
For ethylene, differences between the new cross section fits and the ones stored in HYDKIN have been rather 
small in case of dominating reaction channels, and have been larger for sub dominating ones. However, also in 
the case of dominating channels larger deviations have been observed at low energies (from threshold to 
maximum) reflecting a steeper increase of the cross sections right beyond the treshold. Five completely new 
cross section fits have been generated, which are those for the (sub dominating) 
$(CH_2^+ + F^+)$, $H_2^+$, $H^+$, $C_2H_3^{2+}$ and $C_2H^{2+}$ channels. \\
The largest differences between present and earlier cross section fits have occured in case of propene. While 
the parent ion channel does not deviate much from the counterpart in HYDKIN (within $34\%$), the 
$C_3H_5^+$ and $C_3H_3^+$ channels yield a totally different behaviour. While the 
former is actually the most dominating reaction channel, the latter is almost as large in magnitude as the 
parent ion channel. This reflects a totally different ordering of dominating reaction channels in the case of 
propene, which applies also to the sub dominating channels. Numerous new cross section fits without 
appropriate counterpart in the HYDKIN database have been generated, but lack in information of specified 
fragments. \\
The energy dependence of branching ratios has been investigated, and the (pre-)assumption of energy invariance 
has been found to be well based for dominating reaction 
channels within an accuracy of about $20\%$, and to be worse based for subdominating channels, where 
variations of branching ratios up to factors of 6 have been observed. Therefore, dipole allowed transitions 
remain the major underlying mechanisms for dissociative ionization of the three hydrocarbons under 
consideration, but especially for sub dominating reaction channels alternative non-dipole allowed transitions 
might have a more significant influence than expected from earlier data. This might be especially important when 
constructing cross sections for which no experimental data exists. However, the pre-assumption of energy 
invariant branching ratios has proven to be at least sufficient to produce cross sections in the right order 
of magnitude. \\
On balance, it can be concluded that the differences between 
present data fits and earlier data summarized in HYDKIN cross sections may lead only to small modifications 
of the database, since the differences are rather small on the scale of total ionization cross sections. 
How sensitive simulations might be due to such small changes, could be shown by some sensitivity analysis 
proposed in Ref. \cite{sensitivity}, for which HYDKIN also provides the necessary tools. \\
This work shall be concluded with some general remarks. Firstly, the HYDKIN database as well as the analyzed 
experiments are based on hydrocarbons consisting of hydrogen, obviously dedicating their relevance to hydrogen 
plasmas. It remains questionable if such databases and 
results are applicable also to hydrocarbons consisting of hydrogen isotopes. Secondly, cross section 
experiments as well as calculations are done usually for room temperature. 
However, it has been shown that at least for the energy threshold of electron-impact ionization 
there exist both, 
temperature and isotope effects \cite{mate}. Since the energy threshold plays a crucial role for the energy 
behaviour of cross sections, also these two effects might lead to further modifications and corrections of 
the relevant databases for magnetic confinement fusion.

\renewcommand{\baselinestretch}{1.15}\normalsize

\begin{acknowledgements}
This work was partially supported by the Austrian Science Fund FWF under contract P21061, and by the 
European Communities under the Contract of
Associations between Euratom and the Austrian Academy of Sciences,
carried out within the framework of the European Fusion Development
Agreement. The views and opinions herein do not necessarily reflect those of
the European Commission.
\end{acknowledgements}

% If in two-column mode, this environment will change to single-column format so that long equations can be displayed. 


%\begin{widetext}

%$$\mbox{put long equation here}$$

%\end{widetext}

% Figures should be put into the text as floats. 

% Use the graphics or graphicx packages (distributed with LaTeX2e).


%



% Put the label that you will use with \ref{} command in the braces of the \label{} command.

%

% \begin{figure}



% \end{figure}







%

% \begin{table}



% \end{tabular}

% \end{table}


%\begin{acknowledgments}

% Put your acknowledgments here.

%\end{acknowledgments}


%\bibliography{paper.bib}

\end{document}